\documentclass[%
 a4paper,
 reprint,
 amsmath,amssymb,
 showkeys,prb
]{revtex4-2}

\usepackage{graphicx}
\usepackage{dcolumn}
\usepackage{bm}
\usepackage{hyperref}

\usepackage{physics}
\usepackage[dvipsnames]{xcolor}
\usepackage{cleveref}
\usepackage{nicefrac}
\usepackage[activate={true,nocompatibility},final,tracking=true,kerning=true,spacing=true,factor=1100,stretch=10,shrink=10]{microtype}
\usepackage{orcidlink}

\renewcommand{\vec}{\bm}

\renewcommand{\i}{\ensuremath\textrm{i}}
\newcommand{\oo}{\ensuremath\infty}
\newcommand{\e}{\ensuremath\textrm{e}}

\newcommand{\eps}{\ensuremath\eta}
\newcommand{\diff}{\ensuremath\partial}
\newcommand{\stat}{\ensuremath ^{(0)}}
\newcommand{\inv}{\ensuremath ^{-1}}
\newcommand{\T}{\ensuremath^\textsf{T}}

\newcommand{\HH}{\mathcal H}
\newcommand{\LL}{\mathcal L}
\newcommand{\PT}{\mathcal{PT}}

\newcommand{\sinji}{\sin(m\Delta_{ji})}
\newcommand{\nonrecsym}{v^\text{(0)}_\eps} 

\newcommand{\avggamma}{\bar\Gamma}
\newcommand{\excpt}{\gamma_\text c}

\begin{document}

\title{Dynamics of Aligning Active Matter:\\Mapping to a Schrödinger Equation and Exact Diagonalization}

\author{Tara~Steinhöfel~\orcidlink{0009-0002-9105-4225}}%
\author{Horst-Holger~Boltz~\orcidlink{0000-0003-0013-5186}}%
\email{horst-holger.boltz@uni-greifswald.de}
\author{Thomas~Ihle~\orcidlink{0009-0000-5760-7186}}%

\affiliation{Institute of Physics, University of Greifswald, Felix-Hausdorff-Stra{\ss}e~6, 17489 Greifswald}

\date{\today}

\begin{abstract}
    There has been recent interest in the relaxational modes of small-scale fully connected systems of aligning self-propelled particles (Spera et al., Phys. Rev. Lett. {\bf 132}: 078301 (2024)). We revisit the classical connection between Fokker-Planck and Schrödinger equations to address this by means of exact diagonalization, allowing for rigorous analytical insight into the full spectrum. This allows us to extract exact results which we compare to the existing result from linearized statistical field theory. We derive asymptotically correct analytical results that improve upon the prior approximations. We show that this methodology can fruitfully be extended to the case of non-reciprocal interactions which gives rise to a non-Hermitian Schrödinger problem akin to those in open quantum mechanics. While the non-reciprocity can be chosen such as not to alter the stationary distribution, it fundamentally changes the nature of the steady state which we quantify via the entropy production. We discuss the case of low particle numbers as well as the emergence of mean-field dynamics at large numbers.
\end{abstract}


\maketitle

\section{Introduction}

There is a notable difference between classical and quantum statistical physics with respect to deriving time-dependent solutions to the probability functions. While the Fokker-Planck equation and the Schrödinger equation share conceptual similarities, both describing diffusive evolution of a scalar field, the literature reflects a significant asymmetry in the frequency of their direct solutions. This disparity is largely driven by the rigid functional gestalt of the Schrödinger equation; its inherently Hermitian structure permits a suite of spectral and variational methods that remain elusive for the typically non-Hermitian operators found in Fokker-Planck systems. The spectral approach here, a similarity mapping of classical stochastic dynamics to a Schrödinger equation is a well-established textbook technique~\cite{haken1983,risken1984,ho2008,shizgal} that is commonly discussed in one dimension. In particular, there has been heightened interest in the super-symmetric formulation~\cite{parisi1979,bender1981,bernstein1984,marchesoni1988,reimann2002,junker} that allows for the solution of a broad class of models, with applications ranging to the folding of proteins~\cite{polotto} and option pricing in finance~\cite{henry2007}.  A class of non-equilibrium systems subject to very active ongoing research within statistical physics is {\em active matter}~\cite{ramaswamy2010,marchetti2010,bechinger2016,chate2020,tevrugt}. These are, generally speaking, coarse-grained systems in which energy insertion happens at the smallest scale and effects the system dynamics by being used to generate forces or motion. This is in contrast to open or driven systems wherein energy is inserted on the scale of the system size. An important realization is an ensemble of self-propelled particles. Each particle has an internal mechanism that converts stored energy to motion. Inter-particle interactions can lead to the emergence of novel cooperative phenomena such as the onset of collective motion, also called flocking, or motility induced phase-separation.  To our knowledge, the mapping method has not yet been used in the context of active matter  whose effective models allow for an even greater variety in the structure of the Fokker-Planck equation. In particular, active matter systems are not subject of Newton's third axiom and can have non-reciprocal interactions. More generally, this work will serve as a great demonstration that the spectral approach can successfully be extended to higher dimensions for currently discussed systems of interacting identical or equivalent particles. Importantly, this mapping, the transformation to the {\em Liouville normal form}, is applicable for mapping any Sturm-Liouville equation into the form of a Schrödinger equation~\cite{liouville,courant} and not reliant on the underlying motivation of the Fokker-Planck and, thus, the overall technique that we apply is potentially interesting for a large number of physical and non-physical systems~\cite{haken1975}.

The model we effectively consider has previously been named the {\em Brownian mean-field} model~\cite{chavanis2005,chavanis2011,chavanis2014} as it can be considered the overdamped limit of the Hamiltonian mean-field model~\cite{latora1999,latora2001,yamaguchi2004,mukamel2008}. In this context it was studied as a prototypical model for systems with long-ranged interactions, such as self-gravitating particle assemblies. \citeauthor{chavanis2005}~\cite{chavanis2005} did notice the existence of a mapping between the Fokker-Planck equation of the Brownian mean-field model and a Schrödinger equation, but only considered approximate analytical solutions to it by expanding around the minimum of the effective potential. Our work here is motivated from a different direction and, therefore, deviates in essential parts. Crucially, we are less focused on the typical thermodynamical limit of large particle numbers, in which the mean-field (or Vlasov) description is expected to be exact as this is a fully connected model whence the mean-field nomenclature. We will comment on this limit, but most of our work is at finite system sizes of a few or possibly just two particles. Secondly, we are very interested in transient behavior and seek to understand the full dynamics. Finally, we will explicitly consider non-reciprocal interactions, a hallmark of active systems not present in Hamiltonian (or gradient-flow) systems.

As part of our analysis, we revisit the results of \citeauthor{spera2024}~\cite{spera2024} who studied the nematic version of the Brownian mean-field model by means of a linearized statistical field theory approach based on a mean-field version of the Dean-Kawasaki equation~\cite{dean1996,illien2025}. They found that the fluctuations due to the emergent noise from interactions tend to decrease the mass of excitation modes from the steady state, i.e, the time constant of the exponential relaxation is increased and they support their claims by agent-based simulation of small systems of up to $N=16$ particles. However, their renormalization vanishes in the limit of zero external noise,
in disagreement to numerical experiments. This is probably due to the linearization approximation in this theory since it is known from asymptotically correct kinetic theory~\cite{ihle2023,kuersten2025} that the interaction of active particles leads to an effective internal noise that formally arises from quadratic and higher order nonlinear terms in the angular modes. Our results will confirm the previous finding that the interactions renormalizes the diffusive relaxation to higher masses (slower relaxation) and we are able to present asymptotically correct analytical expressions for both extremal regimes of very small and very large interaction.

On the technical side, we employ direct numerical diagonalization of a finite-dimensional matrix representation of the relevant operators in Fourier space. Our goal is to understand the transient aligning dynamics of self-propelled particles on time scales where the adjacencies can be considered to remain unchanged and to reintroduce some technical aspects~\cite{risken1984} to readers from active matter theory where these techniques have so far not been utilized. This allows for an exact assessment of the full relaxation for small particle numbers without any linearization or renormalization procedure.

The spectral approach pursued here makes Haken's program concrete~\cite{haken1975,haken1983}: the eigenvalue structure directly encodes the hierarchy of instabilities, from the flocking transition to the non-equilibrium symmetry breaking at the exceptional point, all within a single unified framework.

This work is structured as follows. After introduction of our model of interest
in \cref{sec:model}, we demonstrate in \cref{sec:kinetic-theory} that the
relevant Fokker-Planck equation can be mapped onto a Schrödinger equation in
imaginary time for an arbitrary amount of particles, with the appearance of a
skew-Hermitian extension in the case of non-reciprocal interactions. We analyze the spectral properties of this Schrödinger problem and lay out how they relate to the time evolution. We then discuss in detail results for reciprocal interactions and how they connect to the work by \citeauthor{spera2024}~\cite{spera2024} as well as for non-reciprocal interactions. By means of careful analysis, we are able to establish the asymptotic behavior. We end with a discussion of this work and an outlook of possible future extensions.
\section{Model}
\label{sec:model}

The starting point of our analysis is a fairly generic Vicsek-like model of self-propelled particles with alignment interactions~\cite{chate2020}. We consider $N$ identical particles, moving in two dimensions at constant speed, i.e., with velocity $v_0 (\cos \theta, \sin\theta) \T$ wherein $\theta$ is the angle of the internal orientation to an arbitrarily chosen reference axis, and subject to an alignment interaction. From symmetry considerations, it is obvious that such an interaction should only depend on the difference of the angles and necessarily has to reflect the $2\pi$-symmetry in these angles. The simplest way of incorporating this are Kuramoto-model-like~\cite{kuramoto,pikovsky,acebron2005} couplings $\sin(m \Delta \theta)$. This still leaves a fundamental modeling choice, the symmetry of the interaction with commonly employed systems being of either polar ($m=1$) or nematic ($m=2$) symmetry. We account for these by means of a integer parameter $m=1,2,\ldots$~\cite{boltz2024}. The microscopic state of
the system is specified by the particle positions $\vec r_i$ and orientations
$\theta_i$, whose equations of motion are given by
\begin{align}
    \dot{\bm r}_i & = v_0 \; (\cos \theta_i, \sin\theta_i) \T \label{eq:eom-position}                                                           \\
    \dot\theta_i  & = \sum_{j\in \Omega_i} \Gamma_{ij} \sin\left(m(\theta_j - \theta_i)\right) + \sqrt{2 D} \; \xi_i \label{eq:eom-orientation}
\end{align}
where the interaction is termed aligning (anti-aligning) if $\Gamma_{ij} > 0$ ($\Gamma_{ij} < 0$),
and reciprocal in the special case that $\Gamma_{ij} = \Gamma_{ji} \equiv \Gamma$.
Further, the orientations are influenced by a microscopic Gaussian white noise $\xi_i$, representing microscopic degrees of freedom not handled explicitly in the equations of motion. That the ``microscopic'' equations of motion are stochastic is a hallmark of the coarse-grained nature of active matter models~\cite{vrugt2025}. In particular, we consider a white orientational noise
\begin{subequations}
    \begin{align}
        \expectationvalue{\xi_i(t)}          & = 0                                \\
        \expectationvalue{\xi_i(t)\xi_j(t')} & = 2 D\; \delta_{ij}\; \delta(t-t')
    \end{align}
\end{subequations}
with uniform amplitude given by the diffusion constant $D$. One could also consider an additional spatial diffusive noise, but a simple scaling consideration reveals that this is only relevant on lengths that are small compared to the hydrodynamic length scale $\ell_\text{diff}^2 \sim D_{\text{spatial}}/D$ wherein $D_{\text{spatial}}$ is the diffusivity of the spatial noise. Thus, the orientational noise is the only relevant noise for genuinely active processes in which the self-propulsion is not overshadowed by a strong Brownian motion. To simplify the notation, we have opted to refer to this orientational diffusivity by $D$ in the following.

The set $\Omega_i$ is the neighborhood of particle $i$ (excluding $i$),
determining its interacting partners. In general, adjacency is not transitive, $i \in \Omega_j$ does not
follow from $j \in \Omega_i$; thus, there are generally both a topological
notion of non-reciprocity, encoded by the sets $\Omega_i$ at time $t$, as well
as a numerical one given by the $\Gamma_{ij}$ (which remain constant with
time) at play.

In active matter, various models exist for the determination of adjacency, most importantly metric (e.g., $i$ and $j$ are neighbors if $\lvert \vec r_i-\vec r_j\rvert \leq R$ with some threshold distance) and topological metric-free (e.g. $\Omega_i$ consists of the $k$ closest particles to $i$) interactions. In this work, we consider the fully connected case where for the neighborhoods one always has $\Omega_i = \{1,\ldots, N\} \setminus \{i\}$. Such frozen
adjacencies might be viewed as the limit of  interaction ranges $R$ comparable
to the system size $L$, or as a treatment of timescales much smaller than
$R/v_0$. Fixing the adjacency topology decouples the positional and orientational degrees of freedom, and the model
reduces to the fully connected XY model described solely by
\begin{align}
    \dot\theta_i & = \sum_{j\neq i} \Gamma_{ij} \sin\left(m(\theta_j - \theta_i)\right) + \sqrt{2 D} \; \xi_i \label{eq:eom-orientation2} \text{.}
\end{align}
This model has been used as a toy model of systems with long-ranged interactions, such as self-gravitating particle ensembles, under the moniker {\em Brownian mean-field model}~\cite{chavanis2005,chavanis2011,chavanis2014}. One way of deriving it, is to consider the overdamped limit of the Hamiltonian mean-field model~\cite{latora1999,latora2001,yamaguchi2004,mukamel2008} with friction. Similar models arise in the context of synchronization~\cite{kuramoto,pikovsky,acebron2005} whence the couplings are often referred to as Kuramoto-model couplings. As the model is fully connected (equivalent to the $d\to \infty$ limit for purposes of the Ginzburg criterion), the thermodynamic limit is described by mean-field theory. Here, we are explicitly and strictly interested in small particle numbers. Additionally, the validity of using frozen adjacencies~\cite{mora201} limits naturally the timescales and shifts the focus from static quantities to {\em dynamic transients}.

Having direct access to the microscopic model of \cref{eq:eom-orientation}, allows for a direct comparison of the later results from statistical theory with agent-based simulations or, more technical, direct integration of the microscopic equations of motion. For the integration of the stochastic differential equation \cref{eq:eom-orientation}, we use an explicit Euler-Maruyama scheme with a sufficiently small time step and consider ensemble averages of observables.
Unless indicated otherwise, our ensembles contain $10^6$ realizations and we choose a time step of $0.01D\inv$. 
We use a Mersenne twister \cite{matsumoto1998} as (pseudo)-random number generator.

The most commonly discussed case are models that display flocking (global polar, ``ferromagnetic'' order) due to the presence of positive polar couplings. We place no restriction on the signs of $\Gamma_{ij}$, allowing for
``anti-ferromagnetic'' as well as mixed-sign interactions~\cite{toner2024}. Anti-aligning interactions lack the allure of flocking as the epitome of collective motion, but have seen considerable interest from a theoretical perspective~\cite{ihle2023,escaff2024,boltz2025,escaff2025}. Recently, it has been found that intelligent active systems that pursue mutual avoidance tend to maximize their relative velocities, implying anti-aligning effective interactions~\cite{bektas2025}.  We define the
average interaction strength as
\begin{equation}\label{eq:avggamma}
    \avggamma = \frac{1}{N(N-1)} \sum_{ij} \Gamma_{ij}
\end{equation}
and explicitly forbid self-interaction by setting $\Gamma_{ii} = 0$. This is constructed to be an intensive parameter. By our definitions, this element is never used in \cref{eq:eom-orientation2}, but this choice will simplify the notation in the following.  We discuss an explicit non-reciprocity in terms of a matrix $\gamma_{ij}$, which encodes for the deviatory elements and is defined such
that
\begin{equation}
    \Gamma_{ij} = \avggamma + \eps \gamma_{ij}\;,
\end{equation}
where $\eps$ serves as a bookkeeping parameter. It follows from \cref{eq:avggamma}, that $\sum_{ij} \gamma_{ij} = 0$. In particular, we note that $\eps=0$ corresponds to a reciprocal system. We stress that our method is not perturbative with respect to $\eps$. All results for non-reciprocal couplings detailed below hold for any non-zero value of $\eps$.

\section{Kinetic Theory}
\label{sec:kinetic-theory}

\subsection{Fokker-Planck Description}

Under the discussed assumptions, a microscopic state of the system is given by the
vector $\vec Z = (\theta_1, \theta_2, \ldots, \theta_N) = (1, 2,
    \ldots, N)$, wherein we introduce a short-hand notation that identifies a particle's internal angle $\theta_i$ with its index $i$. Applying standard stochastic calculus~\cite{haken1975,risken1984},
the Langevin dynamics for the orientation \cref{eq:eom-orientation2} is equivalent to a Fokker-Planck (FP) equation for the $N$-particle probability density function $P_N(\vec Z, t)$.
Reflective of the conservation of total probability, the FP can always be written as a continuity equation involving the probability density current $\vec J$, assuming the form
\begin{equation}\label{eq:fokker-planck}
    \diff_t P_N =  \LL P_N = -\sum_i \diff_i J_i\;,
\end{equation}
where $\cal L$ is the FP operator.
Using the shorthands $\Delta_{ij} = \theta_i - \theta_j$ and $\diff_i = \pdv*{\theta_i}$, explicitly
\begin{equation}
    \LL = -\sum_i \diff_i (\avggamma+\eps\gamma_{ij}) \sinji + D \sum_i \diff_i^2
\end{equation}
and the components of $\vec J$ are seen to be
\begin{equation}
    J_i = \sum_j (\avggamma + \eps \gamma_{ij}) \sin(m\Delta_{ji}) P_N - D\diff_i P_N\;.
\end{equation}

For the reciprocal case $\eps = 0$, the deterministic dynamics of \cref{eq:eom-orientation2} are of a gradient flow type rendering it a passive problem. Then the stationary probability current $\vec J^{(0)}$ vanishes identically, signaling detailed balance~\cite{haken1975}, and the
stationary state is of the usual Boltzmann form, $P_N^{(0)}|_{\eps=0} \propto \exp(-\beta E(\bm Z))$ with the effective dimensionless energy
\begin{equation}\label{eq:effective-energy}
    \beta E(\vec Z) = - \frac{\avggamma}{m D} \sum_{i, j>i} \cos(m\Delta_{ji}) \; \text{.}
\end{equation}

It has been emphasized before~\cite{mihatsch2025} that in models for active particles with reciprocal Kuramoto-alignment a hidden conserved variable, the sum of all angles $S=\frac{1}{N}\sum_{j=1}^N\theta_j$ exists, at least in the limit of zero external noise. As a consequence, for any number of particles $N$, the evolution of $S$ under the influence of external noise performs a simple random walk and is decoupled from the angular differences.

\begin{figure}
    \centering
    \includegraphics[width=\linewidth]{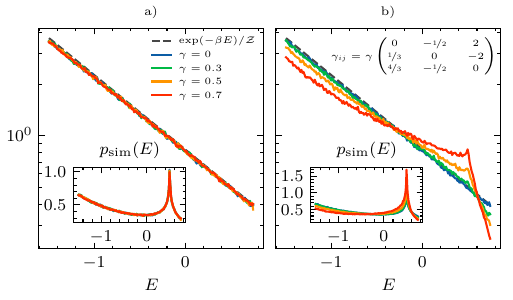}\vspace{-0.25cm}
    \caption{%
        Empirical stationary energy distribution $p_\text{sim}(E)$ as obtained for $\avggamma=0.5$ from agent based simulations. The main panels compare $p_\text{sim}(E)/g_\text{est}(E)$ to the Boltzmann weight $\exp(-\beta E)/\cal Z$, where $g_\text{est}(E)$ is the density of states, estimated by uniform sampling of the configuration space. In the insets we show the empirical density directly without normalization to the density of states. The ensembles consist of $10^6$ systems with $N=3$ particles each, with a runtime of $10D\inv$. \textbf{a)}: Balanced-tournament non-reciprocity as given by \cref{eq:three-particle-nonreciprocity}. The stationary state is Boltzmann distributed, slight tail errors can be attributed to the sampling of $g_\text{est}(E)$. Note that the reciprocal case, $\gamma=0$, is trivially Boltzmann distributed. \textbf{b)}: Same as a), for more arbitrary non-reciprocity as indicated within the panel. The reciprocal energy is no longer a valid potential function in this case.
    }\vspace{-0.25cm}
    \label{fig:stationary-state-is-boltzmann} 
\end{figure}

For non-reciprocal interactions, there will in general be no (analytical) potential function that allows for a direct Boltzmann-like description of the stationary measure. However, we can find a criterion for the Boltzmann weight with the reciprocal energy to still describe the stationary state by plugging it into \cref{eq:fokker-planck} and demanding stationarity. We find a non-vanishing stationary probability current given by
\begin{equation}\label{eq:probability-current}
    J_i\stat = \eps \sum_j \gamma_{ij} \sinji P_N^{(0)} =: v_i\stat P_N\stat
\end{equation}
whose divergence entering the continuity equation \cref{eq:fokker-planck} is
\begin{equation}
    \begin{split} 
        \div J\stat = & \, \eps\sum_{ij} \gamma_{ij} \cos(m\Delta_{ji}) P_N\stat                               \\&+ \eps\sum_{ijk} \gamma_{ij} \frac{\avggamma}{D} \sinji \sin(m\Delta_{ki}) P_N\stat\;.
    \end{split}
\end{equation}
This expression has to vanish for $P_N\stat$ to be a solution of
\cref{eq:fokker-planck}. From the above expression we see immediately a solution for $N=2$ in form of antisymmetric $\gamma_{ij}=-\gamma_{ji}$: the cosine terms are then identically zero due to
the anti-symmetry, while the sine terms cancel identically. For $N=3$ however, antisymmetry does not suffice and the additional condition
\begin{equation}\label{eq:three-particle-nonreciprocity}
    \gamma_{ij} = \begin{pmatrix} 0 & -\gamma & \gamma \\ \gamma & 0 & -\gamma \\ -\gamma & \gamma & 0 \end{pmatrix}_{ij}\;, \quad \gamma \in\mathbb R^+\;,
\end{equation}
has to be met.
While this and the $N=2$ solution are unique (taking the transpose, i.e.\
allowing negative $\gamma$ is equivalent to taking an odd permutation of
particle indices), there is a higher variety of non-reciprocal couplings that do not alter the stationary state for larger values of $N$. Restricting ourselves to the case of couplings of constant strength, i.e., $\lvert \gamma_{ij}\rvert=\text{const}$ for all $i,j$ with $i\neq j$, a nonreciprocal coupling matrix $\gamma_{ij}$ that preserves the reciprocal stationary measure can be identified with  digraphs for which the in- and out-degree are identical. These are commonly called {\em balanced tournaments} in the graph theory literature~\cite{moon2015}. In particular, we are interested in the group of vertex-transitives tournaments, a list of topologically distinct tournaments with $N$ vertices up to $N=49$ is available at ref.~\citenum{mckay}. In this context, the three-particle realization of \cref{eq:three-particle-nonreciprocity}, which corresponds to the structure of the rock-paper-scissors (RPS) game, is the simplest realization and distinct novel graphs exist at odd $N$. Interestingly, there are balanced vertex-transitive tournaments at sufficiently large $N$ for which the corresponding matrix $\gamma_{ij}$ is non-circulant. Here, we use the RPS matrix of \cref{eq:three-particle-nonreciprocity} as an example for which the deviation from standard passive statistical theory induced by non-reciprocal interactions is spurious as the reciprocal stationary measure is still valid.
We demonstrate this in \cref{fig:stationary-state-is-boltzmann}: While for balanced non-reciprocity the stationary measure is Boltzmann distributed, this is not the case for arbitrary $\gamma_{ij}$. The stationary measures were obtained by direct Euler-Maruyama integration of the equations of motion \cref{eq:eom-orientation} and performing an ensemble average.

In general, a Boltzmann-like stationary measure (possibly different from the reciprocal one) exists if the dynamics allow for an orthogonal Helmholtz-Hodge decomposition~\cite{boltz2025}. Throughout this work, we focus on non-reciprocities whose stationary measure corresponds to the reciprocal case. The methodology is suited to treat arbitrary non-reciprocities, but a generic interaction structure will, however, render particles non-equivalent. In the following, we assume equivalent particles that allow for a description of the system in terms of observables that can be measured at any (or all) particles.

\subsection{Schrödinger Description}
\label{sec:schroedinger-desc}

The mapping of a FP equation (or any Sturm-Liouville equation for that matter) to a Schrödinger equation in imaginary time is a standard method of mathematical physics~\cite{liouville,courant,risken1984}. One way of understanding it, is to think about the spectrum of the FP operator. We expect there to be a single stationary ($\lambda\equiv 0$) state  to which all excited eigenmodes relax ($\lambda <0$). If we know the full (discrete, due to the boundary conditions) set of eigenvalues and corresponding eigenmodes, we can describe any transient dynamics in the $N$-particle dynamics exactly. Getting there is more involved from a technical perspective as the FP operator is {\em not} self-adjoint. The knowledge of the stationary state and the resulting symmetric form motivates the representation $P_N = \e^{-\beta E/2} \psi$ as $\partial_t P_N =\e^{-\beta E/2} \partial_t \psi $.
The equation of motion for $\psi$ is then found to be
\begin{equation}\label{eq:fp-operator-conjugation}
    \begin{split}
        -\diff_t \psi & = -\e^{\beta E/2} \LL \e^{-\beta E/2} \;\psi =: \cal H \psi\quad.
    \end{split}
\end{equation}
with a Hamiltonian operator $\HH$.  For brevity we will denote this transformation operator by $\mathcal X$, such that \cref{eq:fp-operator-conjugation} becomes a similarity transform: $-\mathcal X \LL \mathcal X\inv = \HH$ (as it represents multiplication with a non-zero function in real-space, $\mathcal X$ is invertible).  In passive systems with stationary states displaying detailed balance, the term Hamiltonian is justified as $\HH$ is a self-adjoint (Hermitian) operator. As we break this assumption, our operator $\HH$ contains non-Hermitian terms proportional to $\eps$, shown explicitly below. We still stick to the terminology since non-Hermitian Hamiltonians (or Lindbladians) also emerge in genuinely quantum mechanical effective models of open systems with loss or gain \cite{Miri2019,wiersig}.
There, the imaginary parts of the eigenvalues describe the characteristic inverse time scales of energy dissipation.
For us, working in imaginary time, dissipation is the generic feature, and we anticipate the dynamics to acquire oscillatory character in regimes where $\HH$ has complex eigenvalues~\cite{marchetti2010,fruchart2021}. The emergence of non-Hermitian operators in statistical theory is a hallmark of active systems and can generally be related to the broken $\PT$   symmetry~\cite{suchanek2023a,fruchart2021,boltz2026}.

Using the specific $\LL$ of \cref{eq:fokker-planck} the similarity transform of \cref{eq:fp-operator-conjugation}
leads to (details are given in \cref{app:derivation})
\begin{subequations}
    \begin{equation}
        \HH = - D \nabla^2 + V_\text{eff} + \vec{\nonrecsym} \cdot \nabla\;\text{.}
    \end{equation}
    For $\eps=0$, the equation of motion
    \eqref{eq:fp-operator-conjugation} for $\psi$ is thus a Schrödinger equation in
    imaginary time with the effective trigonometric potential 
    \begin{equation}
        \begin{split}
            V_\text{eff} =& \sum_{i, j>i} \bigg[ -m\avggamma \cos(m\Delta_{ji}) \\
            &+  \frac{\avggamma}{2D} (\avggamma + 2\eps\gamma_{ij}) \sinji \sum_k \sin(m\Delta_{ki}) \bigg] \;\text{.}
        \end{split}
        \end{equation}
    For $\eps>0$, we find an additional skew-Hermitian contribution proportional to $\eps$, 
    \begin{equation}
            \vec{\nonrecsym} \cdot \nabla = \eps \sum_{ij} \gamma_{ij} \sinji \; \diff_i \;\text{.}
    \end{equation}
\end{subequations}

For a general stochastic dynamical system $z(t)$ in (for notational brevity) one dimension with a  reciprocal force $F(z)=-\partial_z V$, one finds~\cite{risken1984} $V_{\text{eff}} = \frac{1}{4} \frac{(\partial_z V)^2}{D}-\frac{\partial_z^2 V}{2}$. This ``superpotential'' (including the variant with reversed signs in the latter part) is the starting point of supersymmetric approaches~\cite{junker}.

If we reduce the description to two particles and take only their difference $\Delta=\theta_2-\theta_1$ into account (but not the aforementioned sum $S=(\theta_1+\theta_2)/2$, which decouples in the reciprocal case), then this Schrödinger equations has the form of a Whittaker-Hill equation~\cite{dlmf2}. There is therefore some analytical insight into the eigenfunctions. However, our numerical procedure below generalizes far easier to higher particle numbers and different couplings. We do take note of the fact that the resulting eigenvalue equation will always have strictly periodic coefficients, i.e., be of the form of Hill's equation~\cite{dlmf2} and we can gather from Floquet theory (as well as physical intuition) that the relevant eigenmodes will also be periodic in the angles.

\subsection{Spectral Properties and Stationary State}

Although the mapping presented above is formally reminiscent of quantum mechanics after a Wick
rotation (substituting $-\i t$ for $t$), important differences arise in the
manner the wave functions are translated into probability densities. Recalling the mapping and introduction of the auxiliary function $\psi$ in \cref{sec:schroedinger-desc}, we see that the probability density associated with an eigenstate $\psi_n$ with $n=0,1,\ldots $ is $P_n(x) = \psi_0 \psi_n(x)$. This is a crucial deviation from standard quantum mechanics where the Born rule would apply and has important implications
for the excitations. While in quantum mechanics, any excited state is a valid
physical state \textit{per se}, here a pure excitation of $\HH$ is not actually
physically realizable. To see this we consider the eigenproblem of the FP
operator again, labelling $\LL P_n = \lambda_n P_n$. Since $\LL$ contains only
derivatives, under periodic boundary conditions one finds
\begin{equation}
    \int \!\dd[N] Z \; \LL P_n = 0 = \lambda_n \int \!\dd[N] Z \; P_n
\end{equation}
and hence that the only eigenstate with non-vanishing weight is the (single)
stationary state with $\lambda_0 = 0$. The $P_n$ with $n>0$ do however contain the relaxational information about the dynamics of the time evolution to the stationary state. By formal solution of
\cref{eq:fp-operator-conjugation}, we see that the time evolution is given by $\exp(-\HH t)$.
Since the stationary state is of general interest, but also of concrete importance to the mapping $\LL \mapsto \HH$, we characterize it further.

\paragraph{Probability Density.}

\begin{figure}[b]
    \centering
    \includegraphics[width=\linewidth]{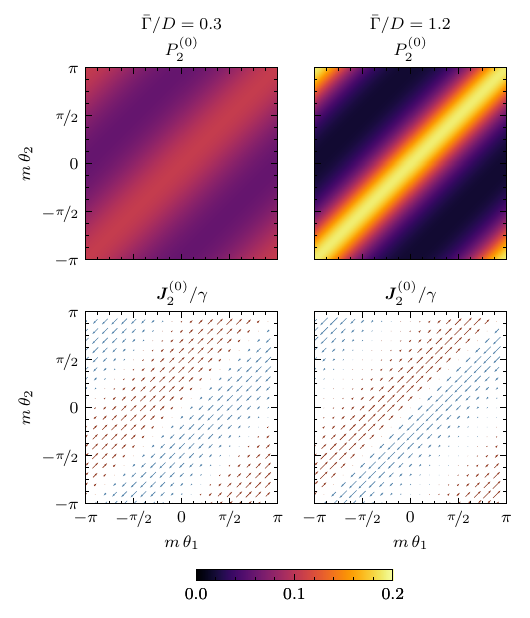}\vspace{-0.25cm}
    \caption{Stationary probability densities as given by \cref{eq:stat-pdf} (top row, heatmap corresponding to the colorbar at the bottom) and currents (bottom row) for $\avggamma/D =0.3$ (left row) and $\avggamma/D=1.2$ (right row) respectively. The periodicity depends on the symmetry number $m=1,2,\ldots$ of the interaction. The currents according to \cref{eq:current} are always along $\bm S$; red and blue colors are used as a guide to the eye to distinguish the sign. For $\gamma=0$,
    interaction is reciprocal and the current vanishes. For the case of negative $\avggamma$, the distributions are shifted relative to the positive $\avggamma$ case by $\pi/m$.
    }
    \label{fig:stationary-current}
\end{figure}

Given reciprocal or globally balanced non-reciprocal interactions as discussed before, direct substitution shows that the stationary probability distribution is (with different normalization) a product of von-Mises distributions
of the $N(N-1)/2$ unique angle differences,
\begin{equation}\label{eq:stat-pdf}
    P_N\stat \propto \prod_{i, j>i} \exp\left(
    \frac{\avggamma}{mD} \cos(m\Delta_{ji})
    \right)
\end{equation}
which is closely approximated by the multi-variate normal distribution, wrapped on a periodic domain.
It is practical to consider a Fourier representation~\cite{dlmf}, also known as the Jacobi-Anger expansion,
\begin{equation}
    \e^{\kappa \cos z} = \sum_{n\in\mathbb Z} I_n(\kappa) \; \e^{\i n z}
\end{equation}
which allows us to relate the Fourier modes of $P_N\stat$ to the modified Bessel functions of the first kind and order $n$, $I_n (\kappa)$,
\begin{equation}\label{eq:jacobi-anger}
    \psi_0 = (P_N\stat)^{\nicefrac{1}{2}} \propto \prod_{\substack{i,j > i}} \sum_{n\in\mathbb Z} I_n(\nicefrac{\kappa}{2}) \; \e^{\i n m \Delta_{ji}}\;,
\end{equation}
where a pre-factor ensuring normalization is omitted and the characteristic value $\kappa$ follows from \cref{eq:stat-pdf} as $\kappa = \avggamma / (m D)$. For $N=2$, an illustration of the stationary density, \cref{eq:stat-pdf}, is given in \cref{fig:stationary-current}.
 
\paragraph{Probability Current.}

While our choice of non-reciprocity as a balanced tournament means that the stationary probability density does not depend on the degree of non-reciprocity $\eps$, the associated non-vanishing probability current shows that
the system resides in a genuine non-equilibrium steady state~\cite{haken1975,seifert2012}. For two particles, introducing $\Delta = \theta_2-\theta_1$ and $S =
    (\theta_1+\theta_2)/2$ as well as the associated direction in phase space, $\vec S = (\hat{\vec \theta}_1 + \hat{\vec \theta}_2)/2$, \cref{eq:probability-current} becomes
\begin{equation}
    \vec J\stat_2 = 2 \gamma \sin(m\Delta) P_2\stat \; \vec{S} \label{eq:current}
\end{equation}
which describes a shear flow in phase space along the $\theta_1=\theta_2$
diagonal. 
This can be observed in \cref{fig:stationary-current}, where we visualize \cref{eq:stat-pdf} and \cref{eq:current} for the case of two particles (for ease of visualization). The relevant normalizing factor is $4\pi^2 I_0(\avggamma/(mD))$ as found by integration using the Jacobi-Anger expansion. The decoupling (in this case of reciprocal interactions) of angle sum and difference is clearly visible. The stationary probability dynamics as indicated by colored arrows in the lower panels is (anti-)aligned with the directions of constant difference.
The probability current can be viewed as a phase space velocity induced by torques on the angle sum
$c$, which together with the equilibrium particle density makes up the
probability current.

\subsection{Time Evolution}

\begin{figure*}
    \centering
    \includegraphics[width=\linewidth]{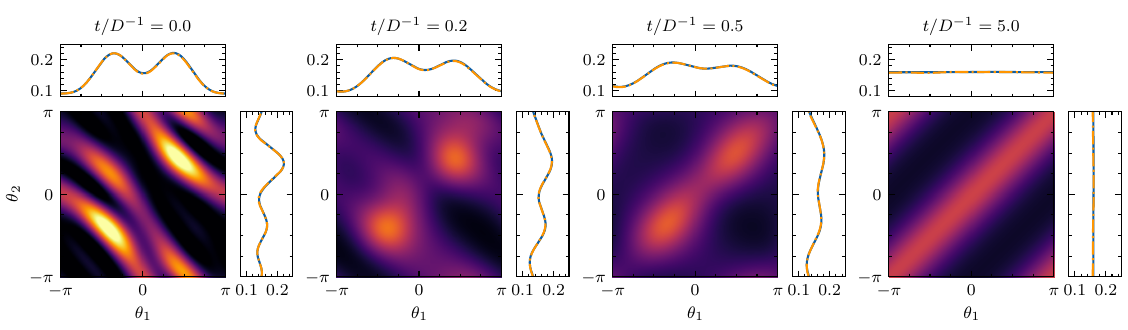}\vspace{-0.25cm}
    \caption{
    Relaxation of an exemplary initial $P_2(\theta_1,\theta_2)$ for $\avggamma/D=1$ and $\gamma_{ij}\equiv 0$. The panels show a time series for $t/D^{-1}=0,0.2,0.5,5.0$.
    Angle-resolved heatmaps where obtained using the mapping method and \cref{eq:time-evolution} (colors as in \cref{fig:stationary-current}).
    Insets show the marginals $P_1(\bm 1)$ (top) and $P_1(\bm 2)$ (right) in blue.
    Reconstructed marginals from empirical Fourier components, obtained from agent-based simulation
    as $\hat p_n^{(i)} = \expectationvalue{\exp(-\i n \theta_i)}$, are drawn in orange dashed-dotted lines,
    showing excellent quantitative agreement.
    In agent-based simulations, the ensemble average $\expectationvalue{\ldots}$ contains $10^6$ systems, the time step is fixed at $0.01/D$.
    The stationary state, despite pronounced correlations, has uniform marginals. The color scheme used here is the same as in \cref{fig:stationary-current}.
    \label{fig:heatmap-time-evolution}
    }\vspace{-0.25cm}
\end{figure*}
The domain of
$P_N$ is an $N$-torus and the initial condition $P_N(t=0)$ is then uniquely represented by a Fourier
series, whose components we summarize in the vector $\vec d(0)$. Formally, the
solution to the time-dependent Schrödinger equation is the propagator
$\exp(-\HH t)$. Given a unitary transformation that diagonalizes $\HH$, such
that $\mathcal U \HH \mathcal U^\dag = \Lambda$, the time evolution of $\vec
    d(0)$ is given by
\begin{equation}\label{eq:time-evolution}
    \vec d(t) = (\mathcal{UX})\inv \e^{-\Lambda t} \mathcal{UX}\;\bm d(0)\;.
\end{equation}

Although in case of a non-Hermitian $\HH$ such a unitary $\mathcal U$ is not guaranteed
to exist, $\PT$ symmetry (of the fully connected XY-model, the original active matter dynamics explicitly break the symmetry) ensures that $\HH$ is diagonalizable except at
isolated spectral singularities, called exceptional points \cite{kato1995}. At
an exceptional point, the $\PT$ symmetry of a given manifold breaks down,
causing these eigenvectors to coalesce, and leaving the Hamiltonian defective.
While in principle the exponential of a defective matrix can still be
efficiently computed from the Jordan canonical form, giving rise to
non-exponential time evolution, the eigenvectors do not perfectly coalesce in
finite precision arithmetic and resulting eigen decompositions are spurious. We
therefore do not consider the dynamics at exceptional points explicitly and remain at finite (but arbitrarily small) distance to them.

The method of choice we employ to study the resulting Schrödinger equation problem is exact diagonalization~\cite{fehske2008}, which in the quantum mechanical literature refers to the direct numerical determination of eigenvalues and eigenvectors from a matrix representation of the problem. When performing exact diagonalization numerically, one can of course only treat finite matrices and, thus, has to introduce a Fourier mode cut-off $\ell$.
Such a cut-off limits the amount of detail which can be resolved in the now truncated Fourier basis. This not only limits which initial conditions can be represented faithfully, but also what interaction strengths can accurately be treated as the stationary state becomes sharply peaked with increasing $\avggamma/D$. The cut-off $\ell$ thus has to be chosen according to this ratio.
In practical calculations, we use the deviation of the lowest lying energy level from zero as a benchmark for the choice of cut-off. 
Including the non-positive Fourier modes, the linear matrix size for a system with $N$ particles is $(2\ell+1)^N$, such that full diagonalization quickly becomes intractable.

For large, sparse non-Hermitian matrices, the Implicitly Restarted Arnoldi~method~\cite{lehoucq1996} is well established as a means of obtaining the extremal, in our case low-lying, parts of the spectrum.
In the case of Hermitian matrices, the Arnoldi method reduces to the Lanczos algorithm \cite{calvetti1994}, which is particularly well-known in the quantum mechanics literature and exact diagonalization context \cite{fehske2008}.
We employ their ARPACK implementations by means of ref.~\cite{haegeman2025}. Parity symmetry can further be exploited to bring the Hamiltonian into block-diagonal form, reducing the effective matrix size by a factor of two.

We show exemplary results of the time evolution in \cref{fig:heatmap-time-evolution} for the case of reciprocal interactions. The different panels represent the two-particle function as a heatmap (cf. fig.~\ref{fig:stationary-current} for details on the representation) at various time steps. We compare the numerical evolution to results from agent-based simulations (i.e. direct numerical integration of the stochastic equations of motion) by means of the marginalized one-particle distributions which are shown in the margins. Not only is an excellent agreement apparent, but it is also clear that the two-particle distribution contains (correlation) information that is not accessible in the marginalized distributions. In this case of reciprocal interactions, the sum and the difference of the two angles decouple. The latter converges to a Von-Mises law whereas the former is subject to free diffusion.

\paragraph{Long-time tail behavior, linearized masses.}

On hydrodynamic scales, the dynamics can be reduced to a small number of slow modes. As we are dealing with identical particles, these are crucially all moments of the one-particle distribution which is the result of marginalizing the degrees of freedom of all but one particle. In our case this reads
\begin{equation}
    P_1(\theta_1) = \int\!\dd{\theta_2}\ldots \dd{\theta_N} \; P_N \;.
\end{equation}
Here, we make use of the physical equivalence of particles that allows us to freely choose the pivot particle.

A suitable representation of the one-particle function are its Fourier modes \begin{equation}
    \hat p_k = \int \!\frac{\dd \theta_1}{2\pi} \; P_1 (\theta_1)\;\e^{-\i k \theta_1} = \frac{1}{2\pi} \expectationvalue{\e^{-\i k \theta_1}} \text{.}
\end{equation}
These not only reflect the symmetry of the angle variable, but also contain the physically relevant information.
The zeroth mode is proportional to the density which is globally conserved and therefore always hydrodynamically relevant and the first mode corresponds to the mean velocity in the active particle setting.
Generally, we would expect the mode with $k=m$, i.e., the mode whose symmetry corresponds to the coupling to be a good order parameter to detect collective behavior. Importantly, the latter definition as an expectation value directly guides their inference from agent-based simulations.

There are eigenmodes in the spectrum of $\HH$ that do not contribute to the marginalized one-particle densities. This emphasizes that a full kinetic approach contains the full hydrodynamic information and, additionally, (correlation) information that is generally inaccessible hydrodynamically.

Close to the stationary state, the effective mode dynamics can be linearized and the modes decay exponentially to their stationary value. In the long time limit, this leads to dynamics that is dominated by the originally excited mode with the slowest decay. Thus, there exists a timescale after which the quantity
\begin{equation}
    \alpha_n (t) = \dv{t} \; {\ln \hat p_n(t)}
\end{equation}
will have become constant, whose exact value (namely one of the entries of $\Lambda^{-1}$ in \cref{eq:time-evolution}) depends on the $n$ of interest.
For ease of comparison to the results of \citeauthor{spera2024}~\cite{spera2024}, who employ an approximate field theory to determine these effective decay rates, we will call the $\alpha_n$ the mass of the $n$th Fourier mode.

\paragraph{Mean Field Result.}

Marginalizing the $N$-particle probability density $P_N$ by integrating out all
degrees of freedom not pertaining to an arbitrary focal particle we label with
the ``$1$'' index, one is left with an equation of the form~\cite{vlasov1938,balescu,kuramoto}
\begin{equation}\label{eq:bbgky}
    \begin{split}
        \pdv{P_1}{t} = & {-}\pdv{\theta_1} (N{-}1)\! \int\!\dd\theta' \;\avggamma \sin(m(\theta'{-}\theta_1)) P_2(\theta_1, \theta') \\& + D\pdv[2]{P_1}{\theta_1}\quad.
    \end{split}
\end{equation}
We note that there are differences in combinatorial prefactors in the context of kinetic theory~\cite{boltz2024}. These works typically deal with dilute systems whereas we are interested in the fully connected case here and keeping track of all factors of $N$ is important. The mean-field (or (McKean-)Vlasov) approximation is based on the assumption of molecular chaos~\cite{kac1956,mckean1967}, $P_2(\theta_1, \theta') =
    P_1(\theta_1) P_1(\theta')$, neglecting correlations between $\theta_1$ and
$\theta'$, thereby closing \cref{eq:bbgky}. Expressing $P_1$ as a Fourier series
with modes denoted by $\hat p_k$, one carries out the
integration over $\theta'$ and finds that the modes couple within the mean-field approximation  like~\cite{kuramoto,pikovsky,acebron2005,peruani2008}
\begin{equation}
    \begin{split}
        \dot{\hat p}_k
         & = k\pi \avggamma (N-1) (\hat p_m \hat p_{k-m} - \hat p_{-m} \hat p_{k+m})  - D k^2 \hat p_k
    \end{split}
\end{equation}
In particular, $\hat p_0 = 1/(2\pi)$ is constant, meaning conservation of probability.
For an interaction of polar (nematic) symmetry, the main mode of interest is
$k=1$ ($k=2$), or generally, $k=m$, which has the equation of motion
\begin{equation}
    \begin{split}
        \dot{\hat p}_m
         & = m \pi \avggamma \left(N-1\right) \left(\frac{\hat p_m}{2\pi} - \hat p_{-m} \hat p_{2m}\right)  - D m^2 \hat p_m .
    \end{split}
\end{equation}
For sufficiently large relaxation times and in disordered states, the term quadratic in the modes will
have become negligible, such that $\dot{\hat p}_m = -\alpha_\text{MF} \hat p_m$
where
\begin{equation}\label{eq:mean-field-mass}
    \alpha_\text{MF} = D m^2 - \frac{m \avggamma}{2} (N-1) 
    \;.
\end{equation}
Notably, this predicts a critical interaction, above which the sign of $\alpha_\text{MF}$ changes and $\hat p_m$ persists, this is the mean-field flocking threshold. For sufficiently large $\bar{\Gamma}$, this seems to imply negative masses, which is an artifact of the then undue linearization, in an order state the modes are not small and the quadratic term cannot be omitted.

We are measuring the masses by studying the relaxation, whereas \citeauthor{spera2024}~\cite{spera2024} considered equilibrium correlation functions. To see that these are indeed equivalent in the case of asymptotically large times (which is the appropriate regime for the linearized dynamics implied by the concepts of single masses), we explicitly show how to compute these generally from the information available within our approach by means of spectral expansion~\cite{risken1984} in \cref{app:eq-corr}.

\section{Reciprocal Interactions}

\begin{figure}
    \centering
    \includegraphics[width=\linewidth]{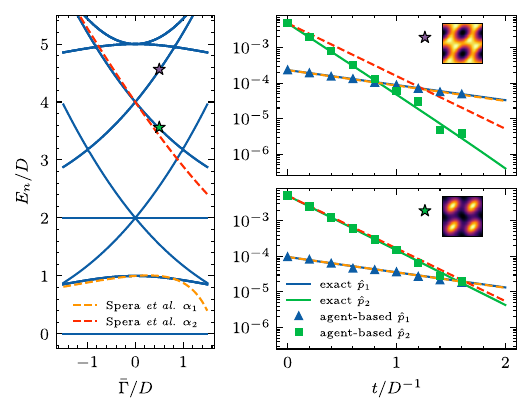}\vspace{-0.25cm}
    \caption{%
        {\bf Left:} Spectrum of $\HH$ for reciprocal interactions $\eps=0$ in the $N=2$ case. We show the eigenvalues $E_n/D$ as a function of the reduced symmetric coupling strength $\avggamma/D$, obtained by exact diagonalization. The flat line $E_0/D \equiv 0$ corresponds to the stationary state. The masses of the polar and nematic fields are given by the eigenvalues of particular eigenmodes, distinguished by symmetry w.r.t.\ particle permutation.
        We consider them exact, since they are the spectrum of $\LL$ by unitary equivalence to $\HH$ and thus entail the full relaxational information. We compare these to the approximate field theory given by \citeauthor{spera2024}~\cite{spera2024} (orange dashed line: polar mode, red dashed line: nematic mode). In between the physically relevant symmetric modes, lie genuine two-particle modes that do not contribute to the order parameter. The two stars mark modes whose relaxation in time is shown in the other panels. There is an apparent deviation of the field theoretic results at larger values of $\avggamma/D$, this is explored in more detail in \cref{fig:projection-approximation-with-spera}.
        {\bf Right:} Relaxation of $\hat p_{1,2}(t)$ for the two eigenmodes which are associated with nematic order, marked in the spectrum with stars, as a function of reduced time. We choose $\avggamma/D = 0.5$ as representative, above this value the deviation of the stochastic theory becomes noticable. The profiles labelled ``exact'' (solid lines) are the marked eigenmodes, relaxing purely exponentially according to \cref{eq:time-evolution}. We find excellent quantitative agreement with direct agent-based simulations (marked by symbols, using a time step of $0.01 D\inv$ and ensemble size of $10^7$) which were initialized to be close to the relevant eigenmodes by inversion sampling from the cumulative distributions corresponding to the eigenmode. In the respective insets, we visualize the modes in the $\theta_1$-$\theta_2$-plane using the same color scheme as in \cref{fig:stationary-current}.
    }\vspace{-0.25cm}
    \label{fig:eigenmodes-spera-comparison}
\end{figure}

In the following two sections, we present explicit results for the relaxational modes and the related masses. Where possible, we compare to analytical asymptotics and approximations. We start with the case of reciprocal interactions, that is a passive fully connected XY model at finite, very small $N$. Indeed, we focus on the case of $N=2$ and increase the system size where needed. At asymptotically large $N$, the mean-field theory sketched in the previous section is exact. It is very important to note here that this method is only limited by the computational effort to diagonalize the relevant matrices and can be applied to any system size. The two-particle dynamics is in that way the most interesting as it is the furthest removed from the regime where mean-field is applicable.
\Cref{fig:eigenmodes-spera-comparison} gives the low-end spectrum for the case of nematic interaction ($m=2$) symmetry. Apart from the spectrum, we show the relaxation of two eigenmodes pertaining to nematic symmetry for the particular choice of $\avggamma/D=\nicefrac{1}{2}$, of which the symmetric one (central panel, of lesser eigenvalue) is associated to the order parameter, and find excellent agreement between the mapping method and agent-based simulations.
The approximate masses from linearized statistical field theory~\cite{spera2024} approximate particular eigenmodes, namely those which originate (in the sense of continuation in $\bar{\Gamma}/D$) from the $\exp(\i m \theta_1)+\exp(\i m \theta_2)$ mode as interaction is turned on. These are the only eigenmodes with respect to which the order parameter $\expectationvalue{\e^{-\i m \theta_k}}$ is both independent of $k$ and non-zero, therefore we refer to them as the principal eigenmodes.
The order parameter is a particular Fourier mode itself, and as such generally contributes to multiple eigenmodes for non-zero coupling $\avggamma/D$ and there is no direct identification between the eigenmodes and the order parameter.
Numerical evidence shows, however, that for e.g.\ $\avggamma/D=2$, the principal eigenmodes' contribution to the order parameter is still over $85\%$. Since this interaction strength is larger than all $\avggamma/D$ considered e.g.\ in \cref{fig:eigenmodes-spera-comparison}, this value acts a lower bound for the ranges considered. This shows that the identified principal eigenmode not only determines the long-time relaxation rates (masses) as argued ealier, but is the dominant contribution to the order parameter.
The related, more general observation that the long-time (or low-energy) behavior of a linear system is well-captured by a small number of dominant eigenmodes underlies a broad 
family of so-called reduced basis methods \cite{christiansen2025}.

The results of \citeauthor{spera2024}~\cite{spera2024} are applicable for nematic, reciprocal interactions and explicitly are using our notations
\begin{subequations}
    \begin{align}\label{eq:spera-polar}
        \alpha_1^{\text{Spera et al.}} & =  D - \frac{4 \avggamma D (N\avggamma - D)}{(5D-N\avggamma) (13 D - N\avggamma)} \\ 
        \alpha_2^{\text{Spera et al.}} & = 4D - N\avggamma + \frac{16 D \avggamma}{(20D-N \avggamma)} \label{eq:spera-nematic} \text{.}
    \end{align}
\end{subequations}
We show these as dashed lines in \cref{fig:eigenmodes-spera-comparison}. 
We highlight that here, $\alpha_m = \alpha_2$ is the mass pertaining to the order parameter. Preempting a result, cf. \cref{eq:projection-approximation} that we will derive in the following section for general (non-reciprocal interactions) by a projection onto the relevant subspace, we give the following small $\avggamma/D$-approximation to the nematic mass
\begin{align} \label{eq:projection_new}
\alpha_{2,\pm} / D = 4 + \frac{\avggamma^2}{4D^2} \pm \avggamma /D\;.
\end{align}

\begin{figure}
    \includegraphics[width=\linewidth]{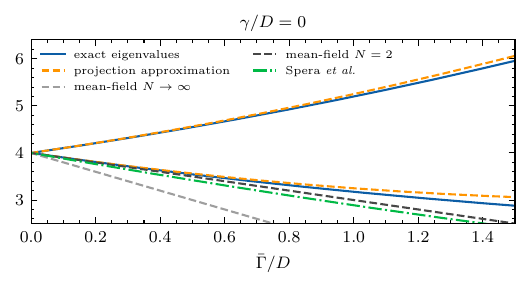}\vspace{-0.25cm}
    \caption{
        Numerically exact eigenvalues (blue solid lines) associated with nematic order in the case of nematic interactions ($m=2$), along with the projective approximation (orange dashed lines) given by \cref{eq:projection_new} and the \citeauthor{spera2024}~\cite{spera2024} approximation (green dash-dotted line) for the nematic mass, as given by \cref{eq:spera-nematic}.
        Also shown are the mean-field results, see \cref{eq:mean-field-mass}, in the thermodynamic limit (light gray dashed line), over which all methods are a considerable improvement, as well as specifically for $N=2$ (dark gray dashed line).
    }\vspace{-0.25cm}
    \label{fig:projection-approximation-with-spera}
\end{figure}

As shown in \cref{fig:projection-approximation-with-spera}, this asymptotically correct approximation is a better incorporation of the fluctuations at small particle numbers  than both the linearized statistical field theory~\cite{spera2024} and the mean-field result. For the extreme case of $N=2$, mean-field actually is an improvement over the field theoretical result. Indeed, the approximate result from linearized field theory is in-between the finite-$N$ and thermodynamical mean-field results for any $N$ and sufficiently small $\avggamma/D$. In this limit, the field-theoretic result expands as  $\alpha_2^{\text{Spera et al.}} \approx 4 D - (N-\nicefrac{4}{5})\avggamma$ and is to be compared with \cref{eq:mean-field-mass}. This highlights an ambiguity in nomenclature as both the resulting theory from assuming molecular chaos factorization (that is of the Vlasov equation) and its thermodynamical (large $N$) limit are commonly referred to as mean-field. As a branch of systematic kinetic theory, mean-field theory does not formally require assuming large system sizes.  Before explaining the methodology to get to the exact results, we also want to highlight that in the complementary limit of large interactions, we can also find exact asymptotic expressions, which we discuss thereafter. 

\begin{figure}
    \centering
    \includegraphics[width=\linewidth]{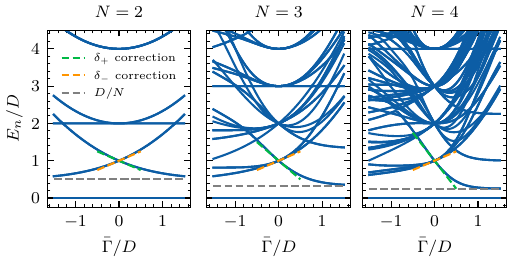}\vspace{-0.25cm}
    \caption{%
        Low-end spectrum for small system sizes and polar alignment interaction ($m=1$), obtained by Lanczos iteration.
        Dashed green (orange) lines indicate the first order correction $\delta_+$ ($\delta_-$).
        Degeneracy is not fully lifted at first order.
        For large $\avggamma$, the low-lying levels are given by $Dn^2/N$. This can be understood by considering a separation of timescales, see main text.
    }\vspace{-0.25cm}
    \label{fig:reciprocal-spectra}
\end{figure}

For weak interactions, we obtain the spectrum perturbatively as corrections to the free problem, $\HH = -D\nabla^2$, via the Hellman-Feynman~\cite{feynman1939} theorem,
\begin{equation}
    \diff_{\avggamma} E_i = \matrixelement{i}{\diff_{\avggamma}\HH}{i}
\end{equation}
where it is implied that the unperturbed eigenstates $\ket i$ diagonalize the perturbation as the perturbation would otherwise diverge.
First order corrections to $E_i$ around the
non-interacting limit are thus obtained from diagonalizing
\begin{equation}
    V^{(1)} =
    \diff_{\avggamma} \mathcal{H} \big{|}_{\avggamma = 0}
    = m \sum_{i, j>i} \cos(m \Delta_{ji})
\end{equation}
in the manifold of interest.
Generally, this is the set of eigenmodes contributing to $\hat p_m$, i.e.\ $\e^{\pm\i m \theta_i}$ in the free limit.
At the degeneracy, this eigenspace is spanned by the basis states
\begin{equation}\label{eq:slowest-eigenspace-span}
    \psi^{(m)}_{j\pm} = \frac{\e^{\pm \i m \theta_j}}{(\sqrt{2\pi})^N}\;, \quad j\in\{1, \ldots, N\} \;.
\end{equation}
For the $m$-manifold, i.e.\ the set of states that become degenerate at $Dm^2$, the matrix representation of $V^{(1)}$ becomes
\begin{equation}\label{eq:effective-hamiltonian-reciprocal}
    [V^{(1)}] = -\frac{m}{2} \left(
    \begin{pmatrix}
            0      & 1      & 1      & \ldots \\
            1      & 0      & 1      & \ldots \\
            1      & 1      & 0      & \ldots \\
            \vdots & \vdots & \vdots & \ddots \\
        \end{pmatrix}
    \oplus
    \begin{pmatrix}
            0      & 1      & 1      & \ldots \\
            1      & 0      & 1      & \ldots \\
            1      & 1      & 0      & \ldots \\
            \vdots & \vdots & \vdots & \ddots \\
        \end{pmatrix}\right)
\end{equation}
where each block corresponds to one sector given by the $\pm$ sign in
\cref{eq:slowest-eigenspace-span} (or, after suitable recombination, the sector
of even and odd eigenfunctions). Each block is an $N\times N$ circulant matrix
generated by the sequence $(0,1,1,1,\ldots)$, and hence we find as our first
order corrections the two eigenvalues
\begin{equation}
    \delta_+ = -\frac{m}{2}\;(N-1) \qq{and}  \delta_- = \frac{m}{2}\;,
\end{equation}
with total multiplicities $2$ and $2(N-1)$, respectively. The subscript indicates whether the related modes are symmetric ($+$) or anti-symmetric ($-$) under exchange of particles. We have confirmed these corrections by exact diagonalization of the full
problem for system sizes up to $N = 7$ particles. Recalling the mean-field
result \cref{eq:mean-field-mass}, one finds
\begin{equation}
    E_m = m^2 +  \frac{\delta_+\avggamma}{D} = \alpha_\text{MF} \;,
\end{equation}
in other words, that the correction $\delta_+$ of bounded multiplicity
reproduces the mean-field result. This highlights that our perturbative approach holds at arbitrary particle number.

The difference in sign of the corrections $\delta_+$ and $\delta_-$ means that
for weak aligning interactions, the $E_m$ eigenspace separates into two
components, such that the lifetime of excitations belonging to the $\delta_+$
($\delta_-$) subspace is enhanced (reduced). From the theory of circulant
matrices the eigenvector to $\delta_+$ is found to be $(1,1,1,\ldots)$, which
represents $\sum_i \cos(\theta_i)$ and $\sum_i \sin(\theta_i)$ in a parity
adapted basis. The qualitative difference in these and the shorter-lived
excitations from their orthogonal complements is symmetry under particle permutation,
which is also explicitly assumed in the derivation of the mean field result.
As the stationary state is symmetric in particle indices, it follows that non-symmetric eigenmodes are unfavorable.

For large positive couplings, $\bar{\Gamma}\gg D$, the diffusive operators act as singular perturbation. While this could be treated by means of the Wentzel-Kramers-Brillouin (WKB) method akin to the quasi-classical limit $\hbar \to 0$ in textbook quantum mechanics, it suffices here to use physical insight to understand the asymptotics of the spectrum, as observable in \cref{fig:reciprocal-spectra}. Going back to the original Langevin equations, there is a clear time-scale separation between the angular differences $\Delta_{1j}$ which will align on some time $\mathcal{O}(\Gamma^{-1})$ and the purely diffusive angular sum $S$ which relaxes on some time $\mathcal{O}(D^{-1})$. We thus can identify $S$ as the slow mode to which the entire system is enslaved. This time-scale separation and the resulting reduction to a single slow degree of freedom is precisely the slaving principle of synergetics~\cite{haken1975,haken1983}: the fast relaxing angular differences are slaved to the slow diffusive mode $S$, which acts as the order parameter governing the macroscopic dynamics.

It is therefore sufficient to determine the relevant Hamiltonian describing $S$.
Since it is purely diffusive, it can be obtained from separation of the Fokker-Planck equation.
Introducing the new set of phase space coordinates,
\begin{equation}
    S = \frac1{N} \sum_{i}^{N} \theta_i \;,
    \quad
    \Delta_j = \theta_1 - \theta_j\; , \quad (1<j\leq N)
\end{equation}
(note that this transformation has unit Jacobian) we split the Laplacian using
\begin{equation}
    \begin{split}
        \diff_1 & = \frac{1}{N} \pdv{S} + \sum_{j=2}^N \pdv{\Delta_{1j}} \\
        \diff_k & = \frac{1}{N} \pdv{S} - \pdv{\Delta_{1k}}\;,\quad k>1
    \end{split}
\end{equation}
such that the pre-factor multiplying $\pdv[2]{S}$ in
$D \nabla^2 = \partial^2_1 + \sum_{k=2}^N \partial_k^2$
is $D/N$. Since the remainder of the Hamiltonian is $S$-independent, it too separates and we find a tower of eigenvalues $Dn^2/N$ for our low-lying levels at large $\avggamma$, which indeed are found in \cref{fig:reciprocal-spectra}. In principle, this insight also holds for \textit{anti-aligning} couplings, but it is less evident, as geometric frustration (the $N=3$ case is identical to the triangular lattice unit cell) will lead to residual energy in the difference angles.

\begin{figure}
    \centering
    \includegraphics[width=\linewidth]{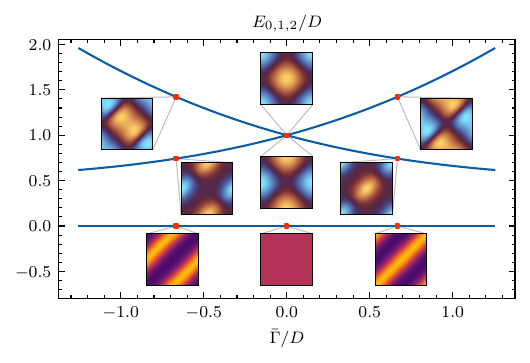}\vspace{-0.25cm}
    \caption{
        Stationary state and splitting of the $E_1$ eigenspace for finite $\avggamma$ and $N=2$. The excited energy levels are two-fold degenerate, shown are pictoral representations of the excitations from the even sector in the fundamental domain of the $(\theta_1,\theta_2)$-plane.
        The excitations re-distribute the weight of the stationary state, shades of yellow (blue) indicate local accumulation (rarification). For comparison, we also explicitly show the ground state $E_0$ here which uses the color scheme of \cref{fig:stationary-current}. In our calculations, we use the deviation of the numerical value of $E_0$ from its exact value $E_0=0$ as a gauge for the numerical precision.
    }\vspace{-0.25cm}
    \label{fig:excitations-with-densities}
\end{figure}

\section{Non-Reciprocal Interactions}

\begin{figure}
    \centering
    \includegraphics[width=\linewidth]{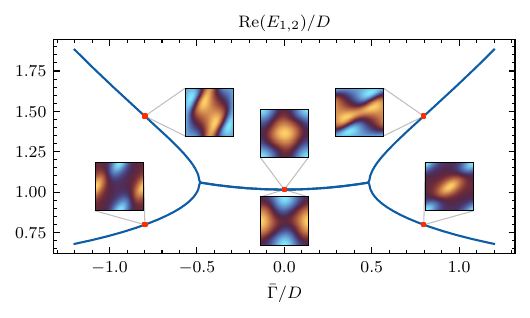}\vspace{-0.25cm}
    \caption{%
        Analogous to \cref{fig:excitations-with-densities}, for finite non-reciprocity $\gamma = 0.5$. As the eigenvalues are generally complex in this case, we only show the real-part which in our setting corresponds to the relaxation. The spectrum exhibits a pitchfork bifurcation at the exceptional point $\avggamma \approx \gamma$.
        Between the exceptional points, $\PT$ symmetry is broken and the eigenvalues are complex-conjugate with equal real part, the non-zero imaginary parts are not shown. 
    }\vspace{-0.25cm}
    \label{fig:nonrec-excitations-with-densities}
\end{figure}

The introduction of non-reciprocity, apart from its implications for the stationary state discused under \cref{sec:kinetic-theory}, allows for the discussion of a wider variety of interactions: on one hand weakly non-reciprocal, ``same-sign'' interactions, where $\gamma_{ij} < \avggamma$, but also ``mixed-sign'' interactions where, conversely, interactions are not consistently aligning or anti-aligning, but lead to chasing motifs~\cite{marchetti2010,fruchart2021,mihatsch2025} due to the competition of the different alignment-preferences of the interacting partners.

On the level of the mapping, this is reflected in the Hamiltonian by the non-Hermiticity allowing for and generally leading to the emergence of an exceptional point, separating two regimes of either broken or intact $\PT$ symmetry. Again, this is in reference to the reduced effective XY-description. In particular, for only weakly non-reciprocal interactions the spectrum remains real (as is ensured in the fully reciprocal case by Hermiticity), whereas for strongly non-reciprocal interactions eigenvalues appear in complex-conjugate pairs (because $\HH$ is purely real).

We will focus on the minimal case of $N=2$, mostly because non-reciprocity is then uniquely parameterized by a single parameter as (with $\gamma>0$)
\begin{equation}
    \gamma_{ij}\big|_{N=2} = \gamma \begin{pmatrix}
        0 & 1 \\ -1 & 0
    \end{pmatrix}\;\text.
\end{equation}
As discussed before, higher particle numbers allow for a great variety of non-reciprocal matrices that preserve the reciprocal stationary measure and, of course, an even larger number that do not.

\begin{figure}
    \centering
    \includegraphics[width=\linewidth]{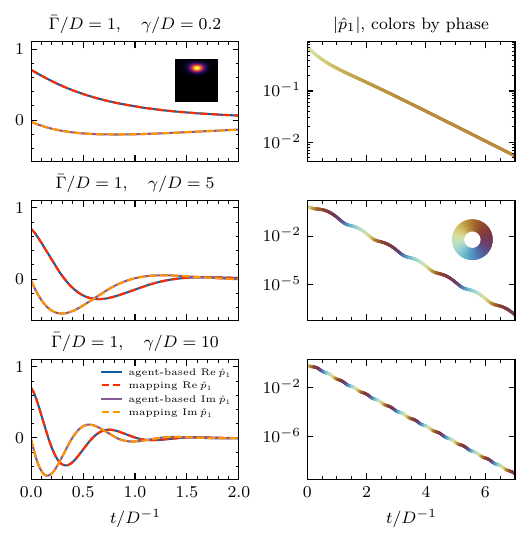}
    \caption{
        Qualitative change in the relaxation of $\hat p_1$ ($m=1$),
        for sub-critical non-reciprocity (first row),
        and non-reciprocity above the exceptional point.
        \textbf{Left:} Comparison of agent-based simulations to the mapping method for the short term, showing quantitative agreement.
        The inset shows the initial condition in the form of a heatmap, using the same color scheme as \cref{fig:stationary-current}. For a symmetric initial condition, $\abs{\hat p_1}$ has zero crossings instead of spiraling behaviour. Solid lines correspond to agent-based results, while the analytical results are shown with dashed lines (the colors are the same throughout and indicated in the lowermost panel).
        \textbf{Right:} Long term evolution of $\abs{\hat p_1}$ using the mapping.
        The envelope is given by $\exp(-\Re E_1 t)$. We again see the emergence of phase (indicated by the coloration of the line) rotation at sufficiently large non-reciprocities $\gamma/D$. 
    }
    \label{fig:time-evolution-non-rec}
\end{figure}

\begin{figure}
    \centering
    \includegraphics[width=\linewidth]{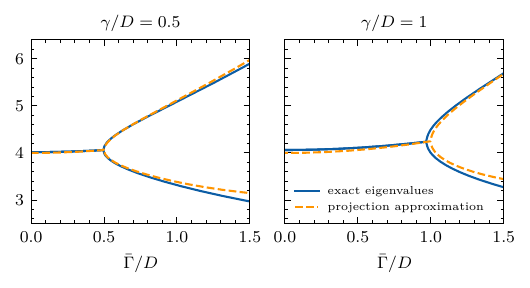}
    \caption{
        Comparison of the projection approximation, \cref{eq:projection-approximation}, to the eigenvalues of the exact problem, here for nematic interactions (see also \cref{fig:projection-approximation-with-spera} for the reciprocal case $\gamma = 0$.)
    }
    \label{fig:projection-approximation}
\end{figure}

For weak interactions, the qualitative behaviour of the spectra seen in \cref{fig:excitations-with-densities,fig:nonrec-excitations-with-densities} can be obtained by projection onto the Fourier modes pertaining to the order parameter.
We define the projector
\begin{equation}
    P_m = \sum_j\ketbra{j+}
\end{equation}
where $\ket{j+}$ represents the basis state given by \cref{eq:slowest-eigenspace-span}.
Thus, we obtain a reduced matrix representation
\begin{equation}
    P_m \HH P_m = \begin{pmatrix}
        Dm^2+\frac{\avggamma^2}{4D}      & -\frac{m}{2}(\avggamma+\gamma) \\
        -\frac{m}{2}(\avggamma - \gamma) & Dm^2+\frac{\avggamma^2}{4D}    \\
    \end{pmatrix}
\end{equation}
which has eigenvalues
\begin{equation}\label{eq:projection-approximation}
    E_{m;1,2} = Dm^2 + \frac{\avggamma^2}{4D} \pm \frac{m}{2} \sqrt{\avggamma^2-\gamma^2}\;.
\end{equation}
This is a good approximation for small $\avggamma$ and $\gamma$, where higher Fourier modes do not couple strongly.
The discriminant predicts the exceptional point to be at $\excpt = \avggamma$ in this case.
While this aligns with the distinction between ``same-sign'' and ``mixed-sign'' interactions given in the beginning of the section,
we numerically find this to only be asymptotically correct for small $\avggamma$, and in general $\gamma_c > \avggamma$.
For example, $\excpt(\avggamma / D=0.5) / D \approx 0.5158(1)$ and $\excpt(\avggamma/D=1) / D \approx 1.1335(1)$ (cf.\ \cref{fig:projection-approximation}).

The relaxational dynamics, shown in \cref{fig:time-evolution-non-rec}, also highlights the clear qualitative change upon crossing the critical level of non-reciprocity. At small non-reciprocity, the relaxation is very similar to the reciprocal case. Upon exceeding the threshold non-reciprocity, an oscillatory behavior (phase rotation) emerges which is reflective of the chasing motif for the two-particle non-reciprocity.

\paragraph{Entropy production} We can quantify  the non-equilibrium stationary state by considering the entropy production corresponding to the non-vanishing probability currents~\cite{seifert2012,fodor2016}.  In general, the entropy production rate in the stationary state can be expressed as
\begin{equation}
    \dot\sigma\stat_N = \frac1{D} \expectationvalue{\vec v\stat \cdot \vec v\stat} = \int \!\dd[N]Z \; \frac{|\vec J_N\stat|^2}{D P\stat_N}\;\text,
\end{equation}
which manifestly vanishes for reciprocal interactions, but is positive otherwise. For two particles, writing the stationary state as $P\stat_N = \exp(\kappa \cos(m\Delta)) / \mathcal N (\kappa)$ where $\mathcal N(\kappa) = 4\pi^2 I_0(\kappa)$ and $\kappa = \avggamma / (Dm)$, this integral can be expressed as
\begin{equation}
    \begin{split}
        \dot\sigma\stat_2 & = \frac{2\gamma^2}{D \mathcal N} \int \sin^2(m\Delta) \; \e^{\kappa \cos(m \Delta)} \; \dd[2] Z \\
                          & = \frac{2\gamma^2}{D \mathcal N} \left(\mathcal N - \pdv[2]{\mathcal N}{\kappa}\right)
        = \frac{\gamma^2}{D} \left(1 - \frac{I_2 (\kappa)}{I_0(\kappa)}\right)
        \;\text.
    \end{split}
\end{equation}
where the identity~\cite{dlmf} $I''_0 = I'_1 = (I_0 + I_2)/2$ has been used.
This is a monotonically decreasing function in $\kappa$ with limits
\begin{equation}
    \lim_{\avggamma\to 0} \dot\sigma\stat_2 = \frac{\gamma^2}{D}
    \qq{and}
    \lim_{\avggamma\to\oo} \dot\sigma\stat_2 = 0
    \;\text.
\end{equation}
The latter in particular means that at fixed $\gamma$, $\avggamma \to \oo$ is asymptotically a reciprocal interaction whose entropy production vanishes. This is an intuitive, yet striking result. In the case of strong alignment interactions, the polar order dominates the system and the active, non-reciprocal structure underneath it is no longer observable. This is in line with other notions that the flocking phase can effectively be modeled as a passive equilibrium state~\cite{fodor2016,mora201,boltz2025}.

\section{Discussion}

Starting with Haken's seminal work, statistical physics has long proven to be a remarkably versatile framework, extending its reach far beyond the confines of traditional thermal systems to describe a vast array of non-equilibrium and, arguably, non-physical phenomena. While this is a valid insight in its own right, it is also bound to highlight the technical challenges of non-equilibrium statistical physics: determination of a non-equilibrium stationary state might be possible in some systems, but access to the full dynamics of the probability density function in interacting systems is remarkably scarce.

We address this mismatch by revisiting a classic rephrasing of Fokker-Planck equations to Schrödinger equations in imaginary time and adapting it to active systems. Two types of activity are relevant for the original context of aligning self-propelled particles that motivated this work. As these particles are moving, their adjacencies are generally subject to change. However, the time-scales on which this happens can be quite large and there are relevant transients on many hierarchy scales. We consider the smallest interacting scale by investigating fixed fully connected networks of small size, corresponding to intermediary clusters. The other type of activity hallmark we explicitly do consider is non-reciprocity of the interactions. Since the interactions in many active systems are effective they are not subject to Newton's third axiom. In general, this leads to descriptions of effective energy landscapes not being applicable~\cite{boltz2024,schüttler2025}, but we show under what conditions the reciprocal energy remains a valid potential function and that the resulting non-Hermitian quantum mechanical problem is still treatable with established techniques of open quantum systems.  As we are considering a fully connected model, textbook statistical physics seems to be rather trivial at first glance. This is a case where the mean-field treatment is exact. However, this statement holds in the thermodynamic limit. Here and in many active matter systems, we are interested in a statistical theory that applies to ensembles of very small systems.

In the reciprocal case, we are able to accurately infer the relaxational time-scales of the order modes which were previously only accessible by means of a uncontrolled linearized statistical field theory~\cite{spera2024}, see \cref{fig:eigenmodes-spera-comparison}. Using the tools developed in genuine quantum mechanical contexts, we are able to derive correct asymptotics and track the transition towards the mean-field result at small to intermediary couplings. We can also derive the asymptotic behavior at large couplings by identifying the angular sum as the slow mode in the system that dominates the dynamics. We are able to qualitatively confirm the previous findings from a solid analytical foundation and can quantitatively improve upon them by providing a rigorous expansion expression.

In the non-reciprocal case, we find that there is a symmetry breaking at a finite level of non-reciprocity manifesting in an exceptional point and oscillatory relaxation corresponding to a chasing motif. This highlights that the mapping to a Schrödinger equation can even be gainful in situations where the resulting quantum mechanical problem is not Hermitian. Furthermore, we have demonstrated under which conditions the reciprocal and non-reciprocal interactions lead to the same stationary distribution and quantified the fundamental difference between the two cases by means of the entropy production.

Going forward, we are hopeful that our result at few-particle scale can be used in meso- and hydrodynamic considerations in the spirit of ref.~\cite {spera2024} as well as guide experimental analysis of the dynamics of dense clusters in flocking systems close but before the onset of flocking (as we are interested in the small-particle limit where deviations from mean-field theory are relevant) and that we have highlighted that the statistical physics machinery is the same for all models and across all scales to the point that techniques from the microscopic quantum world can be applied to ensembles of tangible active objects.

\section*{Data Availability}

Simulation data and source code are available upon reasonable request from the authors.

\appendix


\section{Details of the Mapping}
\label{sec:many-particle-mapping} \label{app:derivation}

While the mapping has been covered in general before~\cite{liouville,courant,risken1984}, we give details of the procedure for our specific model in this section, i.e., we derive
\begin{equation}
    \begin{split}
        \HH & = - \e^{-\beta E/2} \LL \e^{\beta E/2}                        \\
            & = - D \nabla^2 + V_\text{eff} + \vec{\nonrecsym} \cdot \nabla
        \;,
    \end{split}
\end{equation}
where $\abs{\vec{\nonrecsym}}\propto \eps$. The Fokker-Planck operator $\LL$ is of the form
\begin{equation}
    \LL = -\sum_i \diff_i \tilde\alpha_i + D \sum_i \diff^2_i
\end{equation}
where we introduce the shorthands
\begin{align}
    \tilde \alpha_i  & = \alpha_i + \tilde \epsilon_i       \\
    \alpha_i         & = \bar\Gamma \sum_j \sinji           \\
    \tilde\epsilon_i & = \eps \sum_j \gamma_{ij} \sinji \;.
\end{align}
The energy function $E$ is such that its negative gradients give the reciprocal interaction parts, $-\diff_i E = \alpha_i$.
Using the operator identities
\begin{align}
    \diff_i \e^{\beta E/2}   & = \e^{\beta E/2} (\diff_i - \beta \alpha_i/2)   \\
    \diff_i^2 \e^{\beta E/2} & = \e^{\beta E/2} (\diff_i - \beta \alpha_i/2)^2
\end{align}
and sorting orders of $\diff_i$, one finds
\begin{equation}
    \begin{split}
        \HH & = -\e^{-\beta E/2} \LL \e^{\beta E/2}                                                                                                                        \\
            & = \sum_i  \left[ \frac{\alpha_i^2}{4D} + \frac{\alpha_i \tilde\epsilon_i}{2D} + \frac{1}{2} \alpha_i' + \tilde\epsilon_i' + \tilde\epsilon_i \;\diff_i -D\;\diff_i^2 \right]
    \end{split}
\end{equation}
where we used $\beta = 1/D$ and $x'$ denotes multiplication with $\pdv*{x}{\theta_i}$. Expanding the first four terms yields

\begin{widetext}
    \begin{subequations}
        \begin{align}
            V_\text{eff} & = \sum_i \frac{1}{2} (\diff_i\alpha_i) + (\diff_i\tilde\epsilon_i) + \frac{1}{4D} \alpha_i^2 + \frac{1}{2D} \alpha_i\tilde\epsilon_i \\
                         & = - m \sum_{i,j\neq i} \left(\frac{\bar\Gamma}{2} + \eps\gamma_{ij}\right) \cos(m\Delta_{ji})
            + \frac{\bar\Gamma}{2D} \sum_{i,j \neq i,k\neq i}
            \left(\frac{\bar\Gamma}{2} + \eps\gamma_{ij}\right)
            \sin(m\Delta_{ji}) \sin(m\Delta_{ki})
        \end{align}
    \end{subequations}
\end{widetext}
and, thus, the expression given in the main text.

\section{Connection between relaxational dynamics and equilibrium correlators} \label{app:eq-corr}

For the sake of self-containedness, we quickly summarize how our spectral knowledge that contains the full relaxational dynamics can be used to infer equilibrium correlations. This can for example be found in ref.~\citenum{risken1984}, but we use our notation here. As this is a general statement, we keep the presentation general as well.

Suppose we have a Fokker-Planck equation with some Fokker-Planck operator $\mathcal{L}$, i.e. we know that the dynamics of the $N$-particle distribution function $P_N$ is given by
\begin{equation}
    \partial_t P_N = \mathcal{L} P_N
\end{equation}
and we further know the respective eigenvalues $\lambda_i$ as well as the right and left eigenmodes $f_i$ and $g_i$ with
\begin{align}
    \mathcal{L} f_i         & = \lambda_i f_i            \\
    \mathcal{L}^\dagger g_i & = \lambda_i^* g_i \text{.}
\end{align}
The latter is the information we determine by means of the Schrödinger mapping. Although we focus on the right eigenmodes, the left eigenmodes (which generally are not just the complex conjugates of the right modes) are also directly accessible.

The most general equilibrium time correlation function, we could be interested in is of the form
\begin{align}
    C_{AB}(t) & = \langle A(t) B(0) \rangle_{\text{eq}}
\end{align}
with $A$, $B$ being observables that are measured at a time interval $t$ while the system is in equilibrium (hence the subscript). Introducing the conditional probability $P(\vec x,t| \vec x_0, 0)$ of the system to be at the point $\vec x$ in phase-space at time $t$ given that it was at $\vec x_0$ at time $t_0=0$, we can directly write
\begin{align}
    C_{AB}(t) & = \iint_{\vec x, \vec x_0} A(\vec x) P(\vec x,t| \vec x_0, 0) B (\vec x_0) P_N^{(0)}(\vec x_0)
\end{align}
with the equilibrium distribution $P_N^{(0)}$ which we also know. This conditional probability, however, is simply the Green's function (or propagator) of the Fokker-Planck problem. We are thus looking for the solution to the initial condition $P_N(\vec x,0)\propto \delta(\vec x-\vec x_0)$. Using the eigenmodes, we can directly write
\begin{align}
    P(\vec x,t| \vec x_0, 0) =  \sum_i \e^{\lambda_i t} f_i(\vec x) g_i (\vec x_0) \text{.}
\end{align}

Using this expression in the definition of $C_{AB}$, we can directly see that the spectral expansion is indeed directly giving access to the dynamics of the equilibrium correlators as well as
\begin{align}
    C_{AB}(t) & = \sum_i c_i \e^{\lambda_i t} \text{.}
\end{align}
The coefficients $c_i$ here are computable quantities that depend on the modes, the observables and the equilibrium states, but they are also not really needed. If we want to attribute a single time scale to $C_{AB}$, it would have to be the one corresponding to the $\lambda_i$ that is responsible for the slowest decay as this will give rise to a single exponential decay at asymptotically large times.

\nocite{birkhoff,vlasov1968}
\bibliography{references}
\end{document}